\documentclass[12pt]{iopart}
\usepackage[utf8]{inputenc}
\usepackage{graphicx}
\usepackage{amssymb}
\usepackage{aligned-overset}
\usepackage{braket}

\begin{document}

\title[Generalized Bose-Fermi mapping]{Generalized Bose-Fermi mapping and strong coupling ansatz wavefunction for one dimensional strongly interacting spinor quantum gases}

\author{Li Yang\footnote{Now at Google Research, 1600 Amphitheatre Parkway, Mountain View, CA 94043, USA}, Shah Saad Alam \& Han Pu}

\address{Department of Physics and Astronomy, and Rice Center for Quantum Materials, Rice University, Houston, TX 77251, USA}
\vspace{10pt}
\begin{indented}
\item[]June 2022
\end{indented}


\begin{abstract}
Quantum many-body systems in one dimension (1D) exhibit some peculiar properties. In this article, we review some of our work on strongly interacting 1D spinor quantum gas. First, we discuss a generalized Bose-Fermi mapping that maps the charge degrees of freedom to a spinless Fermi gas and the spin degrees of freedom to a spin chain model. This also maps the strongly interacting system into a weakly interacting one, which is amenable for perturbative calculations. Next, based on this mapping, we construct an ansatz wavefunction for the strongly interacting system, using which many physical quantities can be conveniently calculated. We showcase the usage of this ansatz wavefunction by considering the collective excitations and quench dynamics of a harmonically trapped system. 
\end{abstract}

\section{Introduction}
Quantum many-body systems in one dimension (1D) often exhibit unique strongly correlated quantum effects and consequently have attracted much attention over many decades. In recent years, due to their experimental realization in cold atoms, 1D systems have again been at the forefront of active research \cite{Cazalilla2011, Guan2013}. Solving quantum many-body problems, particularly strongly interacting ones, is in general notoriously difficult. This is mainly
due to the fact that there is no general efficient classical computational method to directly solve
these systems, as computational resource required is usually exponential in system
size. However, many powerful analytical (e.g., Bethe ansatz, bosonization) and numerical (e.g., matrix product states) techniques have been developed specifically suitable for 1D systems. Adding to this repertoire, we have recently developed a generalized Bose-Fermi mapping technique that allows us to map a strongly interacting 1D system to a weakly interacting one, which is then amenable for perturbative calculations. This mapping is based on the fact that, in 1D, the distinction between bosons and fermions could become rather subtle, provided
that the bosonic multiple occupancy is suppressed, which can happen if strong repulsion
exists between particles. In this article, we provide a review of this technique and show its application by considering a few examples. 

\section{Generalized Bose-Fermi mapping}
We consider a system of $N$ identical particles of arbitrary spin interacting pairwise via  $s$-wave contact interaction confined in a spin-independent external potential $V(x)$. The Hamiltonian of the system is given by ($\hbar=m=1$)
\begin{equation}
\label{originalHamiltonian}
H=\underbrace{\sum_{i=1}^N\left[-\frac{1}{2}\frac{\partial^2}{\partial x_i^2}+
V(x_i)\right]}_{H_{\rm f}}+\hat{g}\underbrace{\sum_{i<j}\delta(x_i-x_j)}_{V_{{s}}},
\end{equation}
where $H_{\rm f}$ is the single-particle free Hamiltonian and $V_s$ the $s$-wave contact interaction term, with $\hat{g}$ being a matrix acting on the spin state of two particles. There is no constraint on $\hat{g}$ except that it must be symmetric under permutation of two spins so that $V_s$ is invariant under permutation.

For a homogeneous system with $V(x)=0$, Hamiltonian (\ref{originalHamiltonian}) may be Bethe Ansatz solvable. For example, if the system is spinless bosons, this is the Lieb-Liniger model \cite{Lieb1963}; while for spin-1/2 fermions, this is the Gaudin-Yang model \cite{Gaudin1967,Yang1967}. Both models are quantum integrable. In the presence of the inhomogeneous trapping potential, Hamiltonian (\ref{originalHamiltonian}) is in general not analytically solvable. However, Girardeau showed that for spinless bosons in the hardcore limit ($\hat{g}=g \rightarrow \infty$), the system can be solved for arbitrary $V(x)$ by mapping it to free fermions \cite{Girardeau1960}. The eigenstates of the hardcore boson is given by 
\begin{equation}
\label{spinlessInfinite}
\Psi_B(x_1, x_2, ..., x_N) = \sum_{P\in S_N} P(\Psi_F(x_1, x_2, ..., x_N)\theta^1(x_1, x_2, ..., x_N))\,,
\end{equation}
where $P$ is the permutation operator, $\Psi_F$ the free fermion wavefunction, $\theta^1$ is a generalized Heaviside step function of spatial coordinates and can be written into the form:
\begin{align}
\label{theta1}
    \theta^1=\theta(x_{2}-x_{1})\theta(x_{3}-x_{2})\cdots\theta(x_{i}-x_{i-1})\theta(x_{i+1}-x_{i}) \cdots\theta(x_{N}-x_{N-1})\,,
\end{align}
whose value is one in the spatial sector $x_1 < x_2 < ... < x_N$, and zero in any other sector. Eq.~(\ref{spinlessInfinite}) represents Girardeau's Bose-Fermi mapping. It can be easily understood as follows. Within any spatial sector (say the one defined by $\theta^1$), the wavefunction should satisfy the free schr\"{o}dinger equation:
\begin{equation}
\label{freeEq}
    H_{\rm f} \Psi = E \Psi\,,
\end{equation}  
whereas at the boundary of the sector, due to the hardcore condition, the wavefunction should vanish, i.e.,
\begin{equation}
\label{boundaryCd}
    \Psi(x_1, x_2, ..., x_N)_{x_i = x_j} = 0, \;\;\forall \, i, j
\end{equation}
The free fermion wavefunction $\Psi_F$ satisfies both Eq.~(\ref{freeEq}) and the boundary condition (\ref{boundaryCd}). After symmetrization, one arrives at $\Psi_B$ in Eq.~(\ref{spinlessInfinite}) for hardcore bosons. 

We would like to make two important generalizations of the Bose-Fermi mapping:
\begin{enumerate}
    \item Include spin degrees of freedom, hence we can deal with particles of arbitrary spin.
    \item Away from the hardcore limit, i.e., the interaction strength may be finite.
\end{enumerate}

\subsection{Spinor gas with hardcore interaction}
Let us first include the spin degrees of freedom while keeping the interaction in the hardcore limit. The particles are either bosons or fermions with spin-$s$. For this case, we can write the eigenstates in the $\theta^1$ spatial sector in the following form:
\begin{equation}
    \Psi^1 = \varphi^1(x_1, x_2, ..., x_N) \,\chi(\sigma_1, \sigma_2, ..., \sigma_N), \label{scaw1}
\end{equation}
Where $\chi$ is an arbitrary spin wavefunction, such that the spin state is represented by
$\ket{\chi} = \sum_{\sigma_1, \sigma_2, ..., \sigma_N} \chi(\sigma_1, \sigma_2, ..., \sigma_N)
\ket{\sigma_1, \sigma_2, ..., \sigma_N}$, and $\varphi^1 = \varphi \theta^1$ with $\varphi$ being a free fermion eigenfunction (slater determinant) of $H_{\rm f}$. After a symmetrization (for bosons) or antisymmetrization (for fermions), the full wavefunction of the hardcore spinor gas takes the form \cite{Deuretzbacher2008,Guan2009}
\begin{equation}
\label{scaw}
    \Psi = \sum_{P\in S_N} (\pm 1)^P P (\Psi^1) = \sum_{P\in S_N} (\pm 1)^P P \left(  \varphi^1(x_1, x_2, ..., x_N) \,\chi(\sigma_1, \sigma_2, ..., \sigma_N) \right),
\end{equation}
where the permutation operator $P$ is now acting on the indices of both the spatial ($x_i$) and the spin coordinates ($\sigma_i$).

We will call the form of $\Psi$ in Eq.~(\ref{scaw}) the strongly coupling ansatz wavefunction or SCAW. It obviously satisfies the free Schr\"{o}dinger equation (\ref{freeEq}) and the hardcore boundary condition (\ref{boundaryCd}), hence represents the exact wavefunction of the hardcore spinor gas. Two remarks are in order: (i) The SCAW in one spatial sector is a direct product form of spatial and spin wavefunctions (see Eq.~(\ref{scaw1})), but for the full wavefunction (\ref{scaw}), the spatial and the spin degrees of freedom are in general entangled. (ii) Since any spin state will allow Eq.~(\ref{scaw}) to be an eigenstate of the hardcore spinor quantum gas, each eigenstate possesses $(2s+1)^N$ fold degeneracy (ignoring spatial state degeneracy).

\subsection{Spinor gas with finite interaction}
Let us now turn to the case with finite, but still strongly repulsive, interaction. We will show that, to the leading order, the SCAW in Eq.~(\ref{scaw}) remains valid, only that now the spin wavefunction $\chi$ is no longer arbitrary, but is determined by an effective spin chain Hamiltonian. We will proceed by first considering a Hamiltonian duality property for a single particle, followed by a discussion of two interacting particles, and finally the general case of an interacting many-body spinor gas.

\subsubsection{A single-particle Hamiltonian duality ---}
Consider a particle in an arbitrary symmetric potential $V(x) = V(-x)$ with a Dirac  $\delta$-function barrier, governed the Hamiltonian
\begin{equation}
\label{He}
    H^e = -\frac{1}{2}\frac{\partial^2}{\partial x^2} + V(x) + g\delta(x).
\end{equation}
This is a standard textbook problem. Due to the even parity of $H^e$, all its eigenstates possess definite parity. Odd parity states are not affected by the $\delta$-function barrier, hence we just focus on even parity states $\phi(x)$. Integrating the Schr\"{o}dinger equation from $x=0^-$ to $x=0^+$, we obtain
\begin{equation}
\label{boundaryConditionEven}
    \phi^{'}(0^+) = -\phi^{'}(0^-) = g\phi(0),
\end{equation}
where prime denotes derivative with respect to $x$, and the eigen equation on the left and right of the barrier is
\begin{equation}
\label{eigenEquation}
    \left[-\frac{1}{2}\frac{\partial^2}{\partial x^2} + V(x)\right]\phi(x) = E\phi(x).
\end{equation}
With the solution of Eq.~(\ref{eigenEquation}) satisfying the boundary conditon Eq.~(\ref{boundaryConditionEven}), we can obtain all the even eigenstates of Hamiltonian (\ref{He}).

\begin{figure}[ht]
\centering
    \includegraphics[width=14cm]{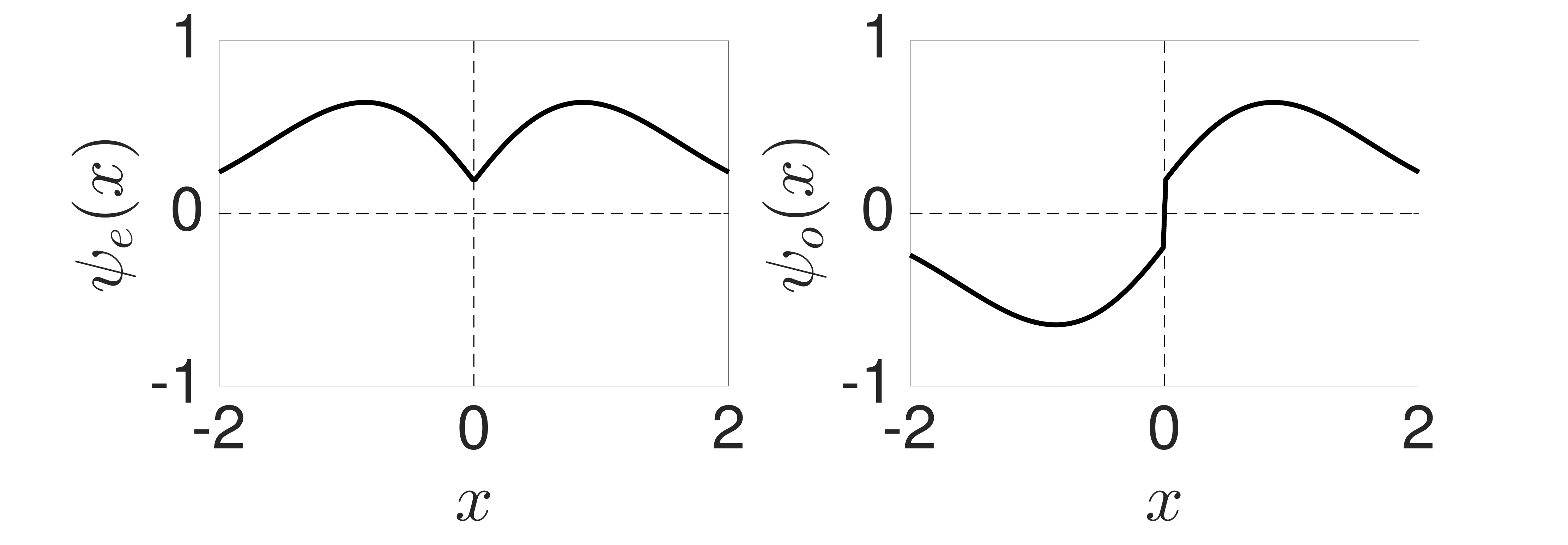}
    \caption{The duality of the $\phi^e$ and $\phi^o$ single-particle wavefunctions, related to each other by Eq.~(\ref{evenOddMapping}). }\label{singleParticleWfDuality}
\end{figure}

Now consider another single-particle Hamiltonian: 
\begin{equation}
\label{Ho}
    H^o = -\frac{1}{2}\frac{\partial^2}{\partial x^2} + V(x)
    - \frac{1}{g}\overset{\leftarrow}{\frac{\partial}{\partial x}}\delta(x)\overset{\rightarrow}{\frac{\partial}{\partial x}}\,,
\end{equation}
where $\overset{\leftarrow}{\frac{\partial}{\partial x}}$ and $\overset{\rightarrow}{\frac{\partial}{\partial x}}$ are differential operators acting on the left and the right wavefunctions when calculating the matrix elements of an operator under a basis, respectively. They are meaningful only when calculating matrix elements of operators. Same as $H^e$, $H^o$ also has parity symmetry. For even states, the $p$-wave singular operator $\overset{\leftarrow}{\frac{\partial}{\partial x}}\delta(x)\overset{\rightarrow}{\frac{\partial}{\partial x}}$ will have no effects, since $\overset{\rightarrow}{\frac{\partial}{\partial x}}$ operators will transform an even state\footnote{Note that the wavefunction is not necessarily continuous. For discontinuous wavefunction, we regard it as a limit of a set of continuous wavefunctions.} to an odd one and the matrix element of $\delta(x)$ will vanish. Let $\phi_m(x)$ and $\phi_n(x)$ be two odd eigenstates of $H^o$ with eigenenergies $E_m$ and $E_n$, and consider the following integral
\begin{equation}
    \int_{0^{-}}^{0^{+}}dx\phi_{m}(x)\left[-\frac{1}{2}\frac{\partial^{2}}{\partial x^{2}}+V(x)
    -\frac{1}{g}\overset{\leftarrow}{\frac{\partial}{\partial x}}\delta(x)\overset{\rightarrow}{\frac{\partial}{\partial x}}\right]\phi_{n}(x)
    =\int_{0^{-}}^{0^{+}}dx\phi_{m}(x)E_{n}\phi_{n}(x)\,.
\end{equation}
Using the fact that $\phi_m(x)$ and $\phi_n(x)$ are odd functions, integrating by parts, we obtain
\begin{equation}
    \phi_{m}(0^{+}) \phi^{'}_{n}(0)
    -\frac{1}{g} \phi^{'}_{m}(0) \phi^{'}_{n}(0) = 0.
\end{equation}
Note that $\phi_{m, n}$ may not be continuous at $x=0$, but since $\phi_n(x)$ is odd, $\phi^{'}_{n}(0)$ is well defined as $\phi^{'}_{n}(0) = \phi^{'}_{n}(0^+) = \phi^{'}_{n}(0^-)$. After factoring out the $\phi^{'}_{n}(0)$ term, we can arrive at a similar boundary condition as
Eq.~(\ref{boundaryConditionEven}),
\begin{equation}
\label{boundaryConditionOdd}
    \phi^{'}(0) = g\phi(0^+) = -g\phi(0^-)\,,
\end{equation}
for any odd eigenstate $\phi$. Comparing Eqs.~(\ref{boundaryConditionEven}) and (\ref{boundaryConditionOdd}), for $x > 0$, the boundary conditions are the same, for $x < 0$, they differ by a sign. This is because we are considering the even eigenstates of $H^e$ and the odd eigenstates of $H^o$, which are dual to each other. And also to the left and right of the $p$-wave singular potential, $\phi_{m, n}$ satisfy the same eigen equation~(\ref{eigenEquation}), which means that the eigenstates and eigenenergies have one-to-one correspondence for $H^e$ and $H^o$ by the relation
\begin{equation}
\label{evenOddMapping}
    \phi^o(x) = \text{sign}(x) \phi^e(x).
\end{equation}
An example is shown in Fig.~\ref{singleParticleWfDuality}.
For odd eigenstates of $H^e$ and even eigenstates of $H^o$, they are trivially dual to each other, and
Eq.~(\ref{evenOddMapping}) still holds. Hence we conclude that the two Hamiltonians $H^e$ in Eq.~(\ref{He}) and $H^o$ in Eq.~(\ref{Ho}) are dual to each other. Due to this duality, we can map the $s$-wave interaction term $g\delta(x)$ to the $p$-wave one $-\frac{1}{g}\overset{\leftarrow}{\frac{\partial}{\partial x}}\delta(x)\overset{\rightarrow}{\frac{\partial}{\partial x}}$.

\subsubsection{Generalized Bose-Fermi mapping for two particles ---} Let us consider two identical particles of arbitrary spin, interacting with contact $s$-wave interaction. The Hamiltonian of the system is given by
\begin{equation}
\label{twoParticleHamiltonianSWave}
    H=\sum_{i=1,2}\left[-\frac{1}{2}\frac{\partial^{2}}{\partial x_{i}^{2}}+V(x_{i})\right]+\hat{g}\delta(x_{1}-x_{2}),
\end{equation}
where $\hat{g}$ is the interaction matrix acting on the spin states of the two particles. By diagonalizing $\hat{g}$, we can fix the spin state $\chi$ to be an eigenstate of $\hat{g}$, which allows us to substitute $\hat{g}$ with the corresponding eigenvalue $g$. And the full wavefunction can be written as
\begin{equation}
    \Psi(x_1, x_2, \sigma_1, \sigma_2) = \Phi(x_1, x_2)\,\chi(\sigma_1, \sigma_2).
\end{equation}
As we have mentioned earlier, $\hat{g}$ must be invariant under permutation. Therefore $\chi$ can have a fixed permutation symmetry, and in turn $\Phi(x_1, x_2)$ should also have a fixed permutation symmetry, since the total wavefunction $\Psi$ must have a fixed permutation symmetry.

For concreteness, let us assume now that the two particles are fermions. To experience the contact $s$-wave interaction, $\Phi$ must be symmetric and $\chi$ then must be anti-symmetric. We can separate $\Phi$ into center-of-mass motion and relative motion:
\begin{equation}
    \Phi(x_1, x_2) = \Phi_c(\frac{x_1 + x_2}{2})\Phi_r(x_{12}),
\end{equation}
where $x_{12}\equiv x_1-x_2$ and the relative motion is governed by the relative Hamiltonian
\begin{equation}
\label{relativeSWave}
    H_r^e = 2\left(-\frac{1}{2}\frac{\partial^2}{\partial x^2_{12}} + \frac{g}{2}\delta(x_{12})\right),
\end{equation}
which, according to the discussion above, is dual to 
\begin{equation}
    H_r^o = 2\left(-\frac{1}{2}\frac{\partial^2}{\partial x^2_{12}} -
    \frac{2}{g}\overset{\leftarrow}{\frac{\partial}{\partial x_{12}}}\delta(x_{12})\overset{\rightarrow}{\frac{\partial}{\partial x_{12}}}\right).
\end{equation}
 Putting things together, we can map the original Hamiltonian~(\ref{twoParticleHamiltonianSWave}) with $s$-wave interaction
to a new Hamiltonian with $p$-wave interaction:
\begin{equation}
\label{_twoParticleHamiltonianPWave}
    H = \sum_{i=1,2}\left[-\frac{1}{2}\frac{\partial^{2}}{\partial x_{i}^{2}}+V(x_{i})\right]
    -\frac{4\hat{P}^{a}}{\hat{g}}\overset{\leftarrow}{\frac{\partial}{\partial x_{12}}}\delta(x_{12})\overset{\rightarrow}{\frac{\partial}{\partial x_{12}}},
\end{equation}
where $\hat{P}^a$ is the projection operator acting on the spin states of the spins, such that $\hat{P}^{a} / \hat{g}$ is nonzero only when the spin state is anti-symmetric. For symmetric spin states, the projections leads to zero. This should be the case, since these states possess anti-symmetric spatial wavefunction and hence do not experience the contact $s$-wave interaction in the original Hamiltonian (\ref{twoParticleHamiltonianSWave}). Hence the new Hamiltonian (\ref{_twoParticleHamiltonianPWave}) is valid for any spin states.

The bases for the Hilbert space of the mapped $p$-wave Hamiltonian~(\ref{_twoParticleHamiltonianPWave}) are
\begin{equation}
\label{twoParticleStatePWave}
    \{\varphi(x_1, x_2) \chi(\sigma_1, \sigma_2)\ | \varphi \in \text{Slater determinants},\chi \in \text{spin states}\}\,,
\end{equation}
where $\{\varphi(x_1, x_2)\}$ is the set of all Slater determinants, and $\{\chi(\sigma_1, \sigma_2)\}$ is the set of spin states without any symmetry constraints. Note that this bases are SCAWs for two particles.

If the two particles are identical bosons, the mapping follows the same derivation as above. The only difference is that, in Eq.~(\ref{_twoParticleHamiltonianPWave}), the anti-symmetric spin projection operator should change to symmetric spin projection operator $\hat{P}^s$.

Finally, we rewrite Hamiltonian~(\ref{_twoParticleHamiltonianPWave}) as
\begin{equation}
\label{twoParticleHamiltonianPWave}
H = \sum_{i=1,2}\left[-\frac{1}{2}\frac{\partial^{2}}{\partial x_{i}^{2}}+V(x_{i})\right]
    -\frac{4\cdot2!\cdot\hat{P}^{s, a}}{\hat{g}}\overset{\leftarrow}{\frac{\partial}{\partial x_{12}}}\delta(x_{12})\theta(x_{12})\overset{\rightarrow}{\frac{\partial}{\partial x_{12}}}.
\end{equation}
We have added $\theta^1$ in the $p$-wave interaction term so that Hamiltonian~(\ref{twoParticleHamiltonianPWave}) is defined in the spatial section with $x_1>x_2$. This is valid since Hamiltonian~(\ref{twoParticleHamiltonianPWave}) acts on the bases of Eq.~(\ref{twoParticleStatePWave}), and the derivative $\partial\varphi({x_{1}, x_{2}}) /\partial x_{12}$ is continuous across $x_{12} = 0$. It is understood that the symmetric (anti-symmetric) spin projection operator $\hat{P}^s$ ($\hat{P}^a$) should be taken for bosons (fermions).

\subsubsection{Generalized Bose-Fermi mapping for many particles ---} Now we can consider a general many-body system consisting of $N$ identical spinful particles. A natural extension of Hamiltonian~(\ref{twoParticleHamiltonianPWave}) to the $N$-body system is given by   
\begin{equation}
\label{pWaveInteraction}
H_p=\underbrace{\sum_{i=1}^N\left[-\frac{1}{2}\frac{\partial^2}{\partial x_i^2}+
V(x_i)\right]}_{H_{\rm f}} \underbrace{- \frac{4N!\hat{P}^{s,a}_i}{\hat{g}}\,\sum_{i=1}^{N-1}\overleftarrow{\partial}_{x_{i,i+1}}\delta(x_{i, i+1})\theta^1\overrightarrow{\partial}_{x_{i,i+1}}}_{V_{{p}}},
\end{equation}
where $\partial_{x_{i,i+1}}=\frac{1}{2}\partial_{x_i}-\frac{1}{2}\partial_{x_{i+1}}$. Hamiltonian~(\ref{pWaveInteraction}) is defined in the $\theta^1$ spatial sector. That we only need to specify the wavefunction in one spatial sector is because of the following. A general $N$-body wave function $\Psi$ for a 1D system can be rewritten as
\begin{equation}
\label{wf}
    \Psi(x_1,x_2,...,x_N,\sigma_1,\sigma_2,...,\sigma_N)=\sum_P (\pm 1)^P P \left(\Psi^1(x_1,x_2,...,x_N,\sigma_1,\sigma_2,...,\sigma_N) \right)\,,
\end{equation} 
where $\Psi^1=\Psi\theta^1$. Equation (\ref{wf}) is a manifestation of a special property of 1D system that the spatial domain of the wavefunction can be separated into $N!$ disconnected subdomains labeled by various spatial orders, and the wavefunction in one spatial sector (say, in spatial sector $\theta^1$) has the complete information of the full wavefunction, as the values of the wavefunction in different spatial sectors are related by permutation operation. The spin projection operator can be written as $\hat{P}^{s,a}_i=(1\pm{\cal E}_{i,i+1})/2$, where ${\cal E}_{ij}$ is the exchange operator that exchanges the $i^{\text{th}}$ and $j^{\text{th}}$ spins. As in the two-particle case, if the original spinor gas is bosonic (fermionic), one should take $\hat{P}^s_i$ ($\hat{P}^a_i$).

The bases for the Hilbert space on which Hamiltonian~(\ref{pWaveInteraction}) operators is given by
\begin{equation}
\label{spinChargeSeparatedBases}
\{\varphi\chi|\varphi\in \text{Slater determinants}, \chi\in \text{spin states}\}.
\end{equation}
That the original Hamiltonian (\ref{originalHamiltonian}) can be mapped to (\ref{pWaveInteraction}) can be understood as follows. The $\delta$-function contact interaction in (\ref{originalHamiltonian}) only introduces the boundary conditions of the eigenstates at spatial sector boundaries. In the region away from those boundaries, the eigenstates are governed by the free Hamiltonian $H_{\rm f}$. The mapped Hamiltonian (\ref{pWaveInteraction}) contains a $p$-wave pseudo interaction potential $V_p$ acting on the Hilbert space defined by (\ref{spinChargeSeparatedBases}), such that its eigenstates, at the boundary of the spatial sector $\theta^1$, are one-to-one mapped to the eigenstates of the original Hamiltonian~(\ref{originalHamiltonian}). As a result, the new Hamiltonian (\ref{pWaveInteraction}) is equivalent to the original Hamiltonian (\ref{originalHamiltonian}), since they possess equivalent eigensystems.

This mapping is valid for any $\hat{g}$. It is particularly useful for a strongly interacting system since it is mapped to a weakly interacting one, with the special case that if the original system has hardcore interaction, the mapped system is non-interacting. Hence our generalized Bose-Fermi mapping contains the Girardeau's Bose-Fermi mapping as a special case. In the following, we will focus our discussion on strongly interacting systems.

\section{Effective spin-chain Hamiltonian and the SCAW}
Now consider a strongly interacting spinor gas governed by Hamiltonian (\ref{originalHamiltonian}). For simplicity we assume that the interaction is spin-independent (i.e., the interaction possesses SU$(2s+1)$ symmetry), or we focus on one particular spin eigenstate of $\hat{g}$, in either case we can replace $\hat{g}$ by a number $g$, which is taken to be large. For more general case where the SU$(2s+1)$ symmetry is broken, a similar approach can be adopted \cite{yangcui2017}. Usually many-body systems with strong interactions are extremely difficult to treat. However, in 1D, as we have shown explicitly in the generalized Bose-Fermi mapping, this is not the case since we can map to the new Hamiltonian $H_p$ in which the interaction term $V_p$ contains a factor $1/g$, hence can be treated as a weak perturbation. Specifically,
working with Hamiltonian $H_p$ in Eq.~(\ref{pWaveInteraction}), the free part $H_{\rm f}$ is considered as the unperturbed Hamiltonian, the interaction part $V_p$ is the perturbing Hamiltonian. We will apply the standard first-order perturbation theory. Since the charge degrees of freedom are described by a spinless Fermi gas, the unperturbed eigenstates are just Slater determinants for free fermions. We label these Slater determinants as $\varphi_n$ with $\varphi_0$ being the ground state, i.e., a filled Fermi sea. We can consider perturbation on any of the unperturbed eigenstates. 

\subsection{Ground-state manifold}
Let us now focus on the ground state. To first order in $V_p$ (i.e., in $1/g$),
we can readily derive an effective Hamiltonian \cite{Yang2015}:
\begin{equation}
\label{heffGround}
    H_{\text{sc}}^{(0)}=E^{(0)}+\langle \varphi_0|V_p|\varphi_0 \rangle = E^{(0)}-\frac{1}{g} \,\sum_{i=1}^{N-1} C_i^{(0)}\,
    (1 \pm {\cal E}_{i,i+1}) \,, 
\end{equation}
where $E^{(0)}$ is the unperturbed ground-state energy, and
the coefficients $C_i^{(0)}$ are given by
\begin{equation}
\label{Ci}
{C}_i^{(0)}=2N!\int dx_1...dx_N \,|\partial_i\varphi_0|^2\,\delta(x_i-x_{i+1})\theta^1 \,.
\end{equation}
Equation (\ref{heffGround}) is an inhomogeneous spin-chain Hamiltonian governing the spin degrees of freedom of the 1D strongly interacting quantum gas. Here the plus (minus) sign should be taken for bosons (fermions). The inhomogeneity stems from the trapping potential $V(x)$, in the absence of which ${C}_i^{(0)}$ become site-independent and we have a homogeneous spin-chain Hamiltonian. The homogeneous spin model is the Sutherland model \cite{Sutherland1975}. Here we want to make two further comments concerning the effective spin-chain Hamiltonian: (1) Note that the coefficients $C_i^{(0)}$, and hence $H_{\text{sc}}^{(0)}$, only depend on the unperturbed Slater determinant $\varphi_0$, which is in turn only dependent on the total number of particles $N$ and the external trapping potential $V(x)$. In particular, $H_{\text{sc}}^{(0)}$ is independent of the spin of the original particles. (2) The spin-chain Hamiltonian is constructed from the nearest-neighbor exchange terms described ${\cal E}_{i,i+1}$. The physics behind this can be intuitively understood as follows: In the hardcore limit $g \rightarrow \infty$, particles in 1D are impenetrable, hence neighboring particles cannot exchange positions. Away from the hardcore limit, the nearest-neighbor exchange becomes possible, and this possibility is captured by the spin-chain Hamiltonian $H_{\text{sc}}^{(0)}$. Our perturbational approach \cite{Yang2015,Yang2016} is inspired by the similar technique used to construct effective spin models from Hubbard Hamiltonian in the large-$U$ limit. Using this technique, the super-exchange interaction arises naturally. Several other groups have obtained the same spin-chain effective Hamiltonian using a variational method \cite{Volosniev2014,Volosniev2015,Deuretzbacher2014,Levinsen2015}.

To leading order, the wavefunction of the system takes the form of the SCAW in Eq.~(\ref{scaw}), where the spin wavefunction $\chi$ is the eigenstate of the spin-chain Hamiltonian. The spin degeneracy for the hardcore system will be (partially) lifted. Let us now take a closer look at this. Consider repulsive\footnote{For attractive interaction with $g<0$, the low-energy states should be bound. Such bound states are not captured by this formalism. However, the unbound states (the so-called upper branch) of the attractive system can still be treated using this approach.} interaction $g>0$. Let us discuss bosons and fermions separately.

For bosons, we need to take the plus sign in the spin-chain Hamiltonian:
\begin{equation}
    H_{\rm boson} = E^{(0)}-\frac{1}{g} \,\sum_{i=1}^{N-1} C_i^{(0)}\,
    (1 + {\cal E}_{i,i+1}) \,.
    \label{hboson}
\end{equation}
Note that coefficients $C_i^{(0)}$ are all positive by definition, see Eq.~(\ref{Ci}), hence the effective spin exchange coupling is ferromagnetic in nature. Each exchange operator ${\cal E}_{i,i+1}$ has eigenvalues $\pm 1$. Hence if we can construct a spin state $\chi_{\rm FS}$ such that
\begin{equation}
    {\cal E}_{i,i+1} |\chi _{\rm FS}\rangle = |\chi_{\rm FS} \rangle\,,\;\; \forall \, i
    \label{fullsym}
\end{equation} 
that would obviously be the ground state of $H_{\rm boson}$. We call such a spin state fully symmetric state, it is not only an eigenstate of $H_{\rm boson}$, but also an eigenstate of all exchange operators ${\cal E}_{i,i+1}$ with the same eigenvalue 1. Such fully symmetric state always exists for any spin configuration. For example, given a spin-1/2 system with two spin-$\uparrow$ and one spin-$\downarrow$ atoms, the fully symmetric state is given by 
\begin{equation}
    |\chi_{\rm FS} \rangle = \frac{1}{\sqrt{3}}\left( | \uparrow \uparrow \downarrow \rangle +| \uparrow \downarrow \uparrow \rangle +| \downarrow \uparrow \uparrow  \rangle \right)
\end{equation}
The corresponding ground-state SCAW for the bosonic system is therefore
\begin{equation}
    \Psi_{\rm boson}=\sum_{P\in S_N} P(\varphi_0 \theta^1 \chi_{\rm FS}) = \left(\sum_{P\in S_N} P(\varphi_0 \theta^1) \right) \otimes \chi_{\rm FS},
    \label{psiboson}
\end{equation}
where we have used the fact that $P(\chi_{\rm FS})=\chi_{\rm FS}$. As a result, the ground state for the bosonic system can be written as a product state of a spatial and a spin wavefunction, each of which is symmetric under permutation. Furthermore, the spatial wavefunction is just the wavefunction of the hardcore spinless bosons.

Now let us consider spinor fermions, for which the effective spin-chain Hamiltonian takes the form
\begin{equation}
    H_{\rm fermion} =E^{(0)} -\frac{1}{g} \,\sum_{i=1}^{N-1} C_i^{(0)}\,
    (1 - {\cal E}_{i,i+1}) \,.
    \label{hfermion}
\end{equation}
Due to the sign change, here the spin exchange coupling is antiferromagnetic.
A similar reasoning as above shows that if we can construct the fully anti-symmetric state such that 
\begin{equation}
    {\cal E}_{i,i+1} |\chi _{\rm FAS}\rangle = -|\chi_{\rm FAS} \rangle\,,\;\; \forall \, i
    \label{fullasym}
\end{equation} 
it will be the ground state of $H_{\rm fermion}$. The corresponding ground state SCAW would be
\begin{equation}
    \Psi_{\rm fermion}=\sum_{P\in S_N} (-1)^P P(\varphi_0 \theta^1 \chi_{\rm FAS}) = \left(\sum_{P\in S_N} P(\varphi_0 \theta^1) \right) \otimes \chi_{\rm FAS},
    \label{psifermion}
\end{equation}
where we have used $(-1)^PP(\chi_{\rm FAS})=\chi_{\rm FAS}$.
Here again the total wavefunction is a product state of a spatial and a spin wavefunction, and the former is again given by the wavefunction of hardcore spinless bosons. However, there is a caveat: the fully anti-symmetric spin state $\chi_{\rm FAM}$ can only be constructed if there is no more than one particle in a given spin state (Hence a necessary condition is that $N\le 2s+1$, i.e., the total number of fermions cannot be more than the spin multiplicity.) \cite{Yang2011}. In the analogous system as considered above: two spin-$\uparrow$ and one spin-$\downarrow$ fermionic atoms, $\chi_{\rm FAM}$ does not exist. In this case, the ground state of $H_{\rm fermion}$ is given by  
\begin{equation}
    |\chi \rangle = \frac{1}{\sqrt{6}}\left( | \uparrow \uparrow \downarrow \rangle -2| \uparrow \downarrow \uparrow \rangle +| \downarrow \uparrow \uparrow  \rangle \right),
\end{equation}
and the corresponding total SCAW cannot be written as a product state of a spatial and a spin wavefunction, indicating entanglement between the spatial and the spin degrees of freedom. Finally, we note that the fully symmetric state $\chi_{\rm FM}$ remains as an eigenstates of $H_{\rm fermion}$ and the associated SCAW is
\begin{equation}
    \Psi_{\rm FS}=\sum_{P\in S_N} (-1)^P P(\varphi_0 \theta^1 \chi_{\rm FS}) = \varphi_0 \otimes \chi_{\rm FS},
\end{equation}
This is again a spin-charge product state and the spatial wavefunction is just the Slater determinant of free fermions. However, this state is not the ground state, and is in fact the highest-lying state in the ground-state manifold.

\subsection{Excited-state manifold}
\begin{figure}[ht]
\centering
    \includegraphics[width=10cm]{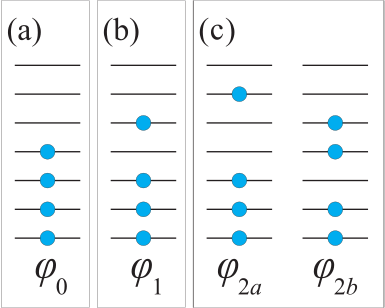}
    \caption{Schematic representation of the ground state (a), the first excited state (b), and the second excited states (c) of an ideal spinless Fermi gas. }\label{2016_fig1}
\end{figure}

In the above, we have focused on the ground-state manifold. Perturbation can be performed on any eigenstates of the unperturbed Hamiltonian, i.e., $H_{\rm f}$. Studies on the excited manifold can provide information on the excitation properties of the system. To show this, let us consider the specific example of a harmonically trapped system with $V(x)=x^2/2$, where we have adopted the natural units system with $\hbar=m=\omega=1$. The ground, first and second excited manifold of a harmonically trapped ideal spinless Fermi gas (corresponding to the eigenstates of $H_{\rm f}$) are schematically shown in Fig.~\ref{2016_fig1}. 

The spin-chain model for the first excited state manifold can be constructed in the similar way as for the ground state manifold. The effective Hamiltonian takes the same form as $H_{\rm sc}^{(0)}$:
\begin{equation}
\label{heffFirst}
    H_{\text{sc}}^{(1)}=E^{(1)}+\langle \varphi_1|V_p|\varphi_1 \rangle = E^{(1)}-\frac{1}{g} \,\sum_{i=1}^{N-1} C_i^{(1)}\,
    (1 \pm {\cal E}_{i,i+1}) \,, 
\end{equation}
where $C_i^{(1)}$ have the same expression as $C_i^{(0)}$ in Eq.~(\ref{Ci}) except that $\varphi_0$ is replaced by $\varphi_1$. For the harmonic trap, the unperturbed eigenenergies are: $E^{(0)}=N^2/2$ and $E^{(n)}=E^{(0)}+n$.

Due to the equal spacing single-particle energy levels for harmonic trap, the second excited-state manifold is doubly degenerate, see Fig.~\ref{2016_fig1}(c). The spin-chain Hamiltonian for the second excited manifold can be written as
\begin{equation}
\label{heffSecond}
    H_{\text{sc}}^{(2)} = E^{(2)}- \frac{1}{g} \,\sum_{i=1}^{N-1} {\textbf{C}}_i^{(2)}\, (1 \pm {\cal E}_{i,i+1}) \,, 
\end{equation} where ${\textbf{C}}_i^{(2)}$ is a $2\times 2$ matrix whose elements are given by 
\begin{equation}
\label{CiUpper}
\left({\textbf{C}}_i^{(2)} \right)_{\alpha \beta} = 2N!\int dx_1...dx_N \, \partial_i\varphi_\alpha \delta(x_i-x_{i+1})\theta^1\partial_{i}\varphi_\beta \,,
\end{equation}
with $\alpha, \beta = 2a, 2b$. Strictly speaking, Hamiltonian~(\ref{heffSecond}) is no longer a pure spin Hamiltonian, as we now have two spatial wave functions $\varphi_{2a,2b}$, which leads to a spin-orbit coupling between the spatial and the spin sectors.

In principle, one can construct the effective Hamiltonian for any other excited manifold in a similar manner as long as we plug in the corresponding Slater determinant(s) to evaluate the coefficients $C_i$. However, due to the special symmetry properties of harmonic trapping potential (specifically the SO(2,1) symmetry \cite{Pitaevskii1997,Werner2006,Nishida2007,Moroz2012}), we can write down the spin-chain model for low-lying excited manifolds from that of the ground-state spin-chain Hamiltonian (\ref{heffGround}) without any extra calculations. The details can be found in Ref.~\cite{Yang2016}. Here we just summarize the main results. By separating the center-of-mass (COM) and the relative motions inside harmonic trap, we can show that the first-excited manifold represents a COM dipole excitation, which are not affected the interaction. Hence we have $C_i^{(1)}=C_i^{(0)}$. As a result, $H_{\text{sc}}^{(1)}$ differs from $H_{\text{sc}}^{(0)}$ by only a constant shift. The doubly degenerate second excited manifold can be separated to two uncoupled modes, denoted as $Q$ and $B$, with associated spin-chain Hamiltonian given by 
\begin{equation}
H^{Q,B}_{\textrm{sc} } = E^{(2)}- \frac{1}{g} \,\sum_{i=1}^{N-1} {C}_i^{(Q,B)}\, (1 \pm {\cal E}_{i,i+1}) \,.
\end{equation}
The $Q$ mode is a COM mode, and for the same reason given above, we have $C_i^{(Q)}=C_i^{(0)}$. The $B$ mode is a relative mode. Quite amazingly, there also exists a simple relation between $C_i^{B}$ and $C_i^{(0)}$ which can be proved using a recursion relation for the SO(2,1) algebra \cite{Levinsen2015,Moroz2012}:
\begin{equation}
\label{factor}
    \frac{ C_i^{B}}{C_i^{(0)}} = 1+ \frac{3}{2(N^2-1)} \,,
\end{equation}
which means that $H_{\textrm{sc} }^{(B)}$ and $H_{\textrm{sc} }^{(0)}$, apart from a constant shift of $E^{(2)}-E^{(0)}=2$, only differ by a constant factor given in Eq.~(\ref{factor}). 

\section{Collective excitations}
The above results provide significant insights into the low-lying collective excitation modes for harmonically trapped spinor quantum gases. That COM modes are not affected by the interaction, but the relative modes are. In particular, let us examine the lowest breathing mode which couples the ground-state to the second excited state manifolds, with the excitation frequency given by
\begin{equation}
\label{ob}
    \omega_B =\langle H_{\textrm{sc} }^{(B)} \rangle -\langle H_{\textrm{sc} }^{(0)} \rangle  = 2 + \frac{3}{2(N^2-1)} E_g \,,
\end{equation}
where $E_g = \langle H_{\textrm{sc} }^{(0)} \rangle - E^{(0)}$ is the ground state energy of the spin-chain Hamiltonian $H_{\textrm{sc} }^{(0)}$ measured with respect to $E^{(0)}$. Hence the breathing mode frequency experiences an interaction-dependent shift away from the non-interacting value of 2. In the strongly interaction regime, this shift $\delta \omega_B \equiv \omega_B-2 \propto 1/g$ and vanishes exactly in the hardcore limit of $g=\infty$. Let us now further examine $\delta \omega_B$ and discuss bosons and fermions separately.

For bosons, as we have discussed above, the ground spin state is the fully symmetric state and
\begin{equation}
\label{eg}
    E_g^{\textrm{boson} }= -\frac{2}{g} \sum_{i=1}^{N-1} C_i^{(0)} \,.
\end{equation}
This result is independent of the spin configuration and only depends on the total number of atoms $N$, a consequence of the fact that the bosonic ground state $\Psi_{\rm boson}$ in Eq.~(\ref{psiboson}) takes the spin-charge separated form. Under the local density approximation (LDA), we can obtain semi-analytic expressions for $C_i^{(0)}$ \cite{Yang2016}, from which, we can show
\begin{equation}
\label{lda}
    E_g^{\textrm{boson} }=-\frac{1}{g}\frac{128\sqrt{2}}{45\pi^2}N^{5/2}\approx -\frac{1}{g}0.408N^{5/2}\,,
\end{equation}
which is consistent with the result obtained previously for spinless bosons near the hardcore limit~\cite{Astrakharchik2005,Zhang2014,Paraan2010}. Correspondingly, the interaction-induced shift of the breathing mode frequency is
\begin{equation}
\label{do}
    \delta \omega_B^{\textrm{boson} } = \frac{3}{2(N^2-1)} E_g^{\textrm{boson} } \approx
    -\frac{1}{g} \frac{64 \sqrt{2}}{15 \pi^2} \,N^{1/2} \,.
\end{equation}

The case for fermions is more complicated.
\begin{itemize}
    \item If the ground-state spin configuration is fully anti-symmetric, i.e., given by $\chi_{\rm FAS}$ with the associated SCAW $\Psi_{\rm fermion}$ given in Eq.~(\ref{psifermion}), then we have $E_g^{\rm fermion}=E_g^{\rm boson}$ and, consequently, 
    \begin{equation}
        \delta \omega_B^{\textrm{fermion} }=\delta \omega_B^{\textrm{boson} }.
    \end{equation}
    However, as we discussed above, the fully anti-symmetric spin state is only possible if no more than 1 fermion occupy one spin component.
    \item For the general case, the ground-state spin configuration is not fully anti-symmetric, and the corresponding SCAW cannot be written as a spin-charge separated form. $E_g^{\rm fermion}$ depends on the specific spin state which is the ground state of $H_{\rm fermion}$. In general, we have $E_g^{\textrm{boson} } \le E_g^{\textrm{fermion} } \le 0$.
\end{itemize}

\begin{figure}[ht]
\centering
    \includegraphics[width=12cm]{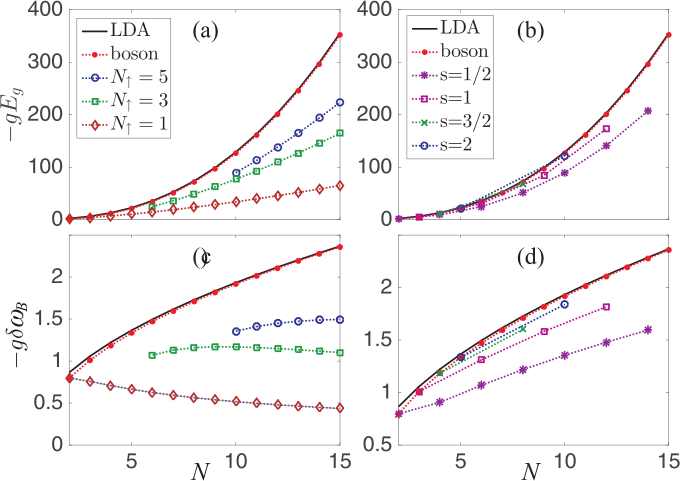}
    \caption{Ground-state energy $E_g$ (a, b) and breathing mode frequency shift $\delta \omega_B$ (c, d) as functions of $N$. In ({a}) and ({c}), we present results for bosons and spin-$1/2$ fermions with various $N_{\uparrow}/N$. In ({b}) and ({d}), we present results for bosons, and fermions with different spin $s$ and equal population in each spin component. For bosons, the ground state energy and the breathing mode frequency shift are independent of spin. The black solid lines represent the analytic LDA results for bosons given in Eqs.~(\ref{lda}) and (\ref{do}). Figure extracted from Ref.~\cite{Yang2016}.}\label{OmegaFig}
\end{figure}

In Fig.~\ref{OmegaFig}(a) and (b), we plot the spin-chain ground state energy $E_g$ as functions of total atom number $N$, with the corresponding breathing mode frequency shift $\delta \omega_B$ plotted in Fig.~\ref{OmegaFig}(c) and (d). The symbols are obtained by numerically calculate the coefficients $C_i^{(0)}$ and then diagonalize the spin-chain Hamiltonian $H_{\rm sc}^{(0)}$. The red dots are the results for bosons. We also plot the analytical results based on LDA (Eqs.~(\ref{lda}) and (\ref{do})) as black solid lines. As one can see, the LDA results agree very well with the numerical results even for small $N$. As one can see, for fixed $N$, as $s$ increases, the fermionic results approach the bosonic ones. This behavior has been recently seen in the experiment \cite{Pagano2014}.

\section{One-body density matrix, momentum distribution and dynamical fermionization}
In this section, we show how the form of SCAW allows us to efficiently evaluate one-body density matrix (OBDM), using which all one-body quantities can be calculated. In particular, we show how to calculate the momentum distribution of a strongly interacting spinor gas. Finally, we discuss the phenomenon of dynamical fermionization.

\subsection{One-body density matrix}
Given a many-body wavefunction $\Psi(x_1,...,x_N;\sigma_1,...,\sigma_N$), the OBDM is defined as
\begin{align}
\begin{split}
\label{OBDM}
\rho(x',x;\sigma',\sigma)
=N\sum_{\sigma_1,...,\sigma_{N-1}}\int dx_1... dx_{N-1} &\Psi^*(x_1,..., x_{N-1}, x';\sigma_1,...,\sigma_{N-1},\sigma') \\
&\times \Psi(x_1,..., x_{N-1},x;\sigma_1,...,\sigma_{N-1},\sigma)\,.
\end{split}
\end{align}
For an SCAW given in Eq.~(\ref{scaw}), the OBDM takes the form
Substituting Eq.~(\ref{scaw}) into Eq.~(\ref{OBDM}), we have
\begin{equation}
    \rho(x',x;\sigma',\sigma)=\sum_{\sigma_1\cdots\sigma_{N-1}}\int dx_1\cdots dx_{N-1}\varphi'^*\varphi\sum_{P'P}\theta'^{P'}\theta^{P}\otimes(P'\chi'^{\dagger})(P\chi)\,,\label{obdm1}
\end{equation}
where we have used the short-hand notation 
\begin{eqnarray*}
&\varphi' = \varphi(x_1,..., x_{N-1},x') \;, \varphi = \varphi(x_1,..., x_{N-1},x) \;, \\
&\chi' = \chi(\sigma_1,...,\sigma_{N-1},\sigma') \,, \chi = \chi(\sigma_1,...,\sigma_{N-1},\sigma) \,.
\end{eqnarray*}
To evaluate the above equation, we need to order $x'$ and $x$ with respect to $x_1,...,x_{N-1}$. For example, assuming $x'<x$, we can take $x' \in (x_{m-1},x_m)$ and $x \in (x_{n-1},x_n)$ with $m\le n$, and denote this ordering configuration as $\Gamma_{m,n}$, in which
    \begin{equation}
        \Gamma_{m,n}:\;\;x_{1}<...<x_{m-1}<x'<x_m<...<x_{n-1}<x<x_n<...<x_{N-1}\,.
    \end{equation}
Once the ordering of $x'$ and $x$ are fixed, all permutations on $1\cdots N-1$ will lead to the same integral value,
because these kind of permutations does not change either $\theta'^{P'}\theta^{P}$ or $(P'\chi'^{\dagger})(P\chi)$ .
According to this observation, the OBDM~(\ref{obdm1}) can be written as \cite{Yang2015,Deuretzbacher2016}
\begin{equation}
    \rho(x',x;\sigma',\sigma)=\sum_{m,n=1}^{N}\rho_{mn}(x',x)S_{mn}(\sigma',\sigma)\,.\label{OBDMspincharge}
\end{equation}
Equation (\ref{OBDMspincharge}) takes a kind of ``spin-charge" separated form, which is a consequence that the SCAW has the spin-charge separated form in any given spatial sector. Here the
spatial part
\begin{equation}
    \begin{split}
        \rho_{mn}(x',x)=&(-1)^{n-m}N!\int_{\Gamma_{m,n}}dx_{1}...dx_{N-1}\,
        \varphi'^{*} \,\varphi \,,\label{SOBDM}
    \end{split}
\end{equation}
depends only on the charge state, i.e., the Slater determinant for non-interacting spinless fermions, $\varphi$, and hence is ``{\it universal}".
The information on the spin degrees of freedom is carried by the spin correlation function
\begin{align}
\begin{split}
\label{Smn}
    S_{mn}(\sigma',\sigma)=(\pm1)^{m - n}\langle{\chi|S_m^{\sigma',\sigma}(m...n)|\chi} \rangle\,, 
\end{split}
\end{align}
(again, $\pm1$ for bosonic and fermionic gases, respectively) where $S_m^{\sigma',\sigma}$ is a local SU($N$) generator
($S^{\sigma',\sigma}\ket{\sigma}=\ket{\sigma'}$) on site $m$, and $(m...n)$ is a loop permutation operator that permutes the indices in the wavefunction by 
$m\rightarrow m+1,m+1\rightarrow m+2,...,n-1\rightarrow n,n\rightarrow m$.
In the above, we have assumed that $m\le n$.
The case with $m\ge n$ can be obtained using the identity $\rho_{mn}(x',x) =\rho_{nm}(x,x')$ and $S_{mn}(\sigma',\sigma)=  S_{nm}(\sigma,\sigma')$.

The difficulty of evaluating the OBDM lies in the fact that Eq.~(\ref{SOBDM}) involves an ($N-1$)-dimensional integral.
With sophisticated numerical techniques, one may be able to carry out such an integral up to $N\sim 20$~\cite{Deuretzbacher2016}.
We have developed a new method \cite{Yang2017} to evaluate $\rho_{m,n}(x',x)$, which relies on its discrete Fourier transform given by:
\begin{equation}
    \rho_{mn}(x',x) = N^{-2} \sum_{\kappa, \kappa'} \rho^{\kappa',\kappa}(x',x) \,e^{i\pi\kappa'm} \,e^{-i\pi\kappa n}\,,
\end{equation}
where $\kappa$ and $\kappa'$ take a discrete set of values $2k/N$ with $N$ consecutive integers $k$, and
    \begin{equation}
        \rho^{\kappa',\kappa}(x',x)=N\int dx_{1}...dx_{N-1}\prod_{j=1}^{N-1}A^{\kappa'*}(x_{j}-x')A^{\kappa}(x_{j}-x) \,\varphi'^{*} \,\varphi \,,\label{rhok1k2}
    \end{equation}
where $A^{\kappa}(x_{i}-x_{j})\equiv e^{i\pi(1-\kappa)\theta(x_{i}-x_{j})}$.
Remarkably,
\begin{equation}
    \Psi^\kappa (x_1,...,x_N) =\left[ \prod_{i<j} A^{\kappa}(x_{j}-x_i) \right] \,\varphi(x_1,...,x_N)\,,\label{psikappa}
\end{equation}
is the wavefunction of $N$ hardcore spinless anyons~\cite{Zhu1996, Girardeau2006} with statistical parameter $\kappa$
(we use the convention in Ref.~\cite{Calabrese2007,Santachiara2007, Santachiara2008}), whose OBDM,
$\rho^\kappa(x',x)\equiv \rho^{\kappa,\kappa}(x',x)$, is given exactly by Eq.~(\ref{rhok1k2}) with $\kappa'=\kappa$. 
The case with $\kappa=0$ and 1 correspond to the hardcore spinless bosons and the ideal spinless fermions, respectively.
By defining a similar Fourier transform for the spin correlation function
\begin{equation}
    S^{\kappa',\kappa}=N^{-2}\sum_{m,n=1}^{N}S_{mn}e^{i\pi\kappa'm}e^{-i\pi\kappa n}\,,
\end{equation} 
we can rewrite
Eq.~(\ref{OBDMspincharge}), the OBDM of a strongly interacting spinor quantum gas, as
\begin{equation}
    \rho(x',x;\sigma',\sigma)=\sum_{\kappa',\kappa}\rho^{\kappa',\kappa}(x',x)S^{\kappa',\kappa}(\sigma',\sigma) \,. \label{mainresult}
\end{equation}
There has been an extensive study of the properties of 1D hard-core spinless anyon gases
~\cite{Zhu1996,Girardeau2006,Calabrese2007,Santachiara2007,Santachiara2008,Kundu1999,Batchelor2006_1, Batchelor2006_2, Batchelor2007_1, Batchelor2007_2,Patu2007, Patu2008_1, Patu2008_2, Patu2009, Patu2010,Rigol2004,Rigol2005,Wright2014,Hao2008, Hao2009,Hao2016, Hao2017,Marmorini2016,Papenbrock2003,Forrester2003}
(and the references therein).
In particular, their OBDM and momentum distributions have been calculated.
We can take advantage of these results to evaluate Eq.~(\ref{mainresult}) in a very efficient way.
In the following, we consider the momentum distribution of a homogeneous system with translational invariance.

\subsection{Momentum distribution}
Given the OBDM $\rho(x',x;\sigma',\sigma)$, the momentum distribution for spin compoment-$\sigma$ can be obtained as 
\begin{equation}
    \rho_\sigma(p) = \frac{1}{2\pi} \int dx \int dx' \, e^{ik(x-x')} \rho(x',x;\sigma,\sigma).
\end{equation}
For a translational invariant system with length $L$ (periodic boundary condition is assumed), the OBDM
$\rho(x',x;\sigma',\sigma)$ depends only on $y\equiv x-x'$, and Eqs.~(\ref{OBDMspincharge}) and (\ref{mainresult}) are
reduced to
\begin{eqnarray}
    \rho(y;\sigma',\sigma) &=& \sum_{r=0}^{N-1} \rho_r(y)\, S_r(\sigma',\sigma) 
    =\sum_{\kappa} \rho^\kappa(y) S^\kappa (\sigma',\sigma) \,,\label{hobdm}
\end{eqnarray}
where $r$ in the first line is understood as $n-m$, so from Eq.~(\ref{Smn}) we have
$S_r(\sigma',\sigma) = (\pm1)^{r}\langle{\chi|S_m^{\sigma',\sigma}(m...m+r)|\chi}\rangle$ which is independent of $m$. To ensure the boundary condition, we need to impose the selection rule $(1...N)\chi=(\mp1)^{N-1}\chi$ on the spin state
$\chi$ with $\mp1$ for bosonic and fermionic gases, respectively.
After Fourier transform with respect to $y$, the corresponding momentum distribution for the spinor quantum gas can be
obtained as
\begin{equation}
    \rho_\sigma(p)=\sum_{\kappa}\rho^{\kappa}(p)\,S^{\kappa}(\sigma,\sigma) \,,\,\label{MomentumTI}
\end{equation}
where $\rho^{\kappa}(p)$ is the momentum distribution for the hardcore anyon system.
Note that $\rho^\kappa$ and $S^\kappa$ are periodic in $\kappa$ with period 2.
Hence we may restrict $\kappa$ in the range $[-1, 1]$.

\begin{figure}[ht]
\centering
\includegraphics[width=12cm]{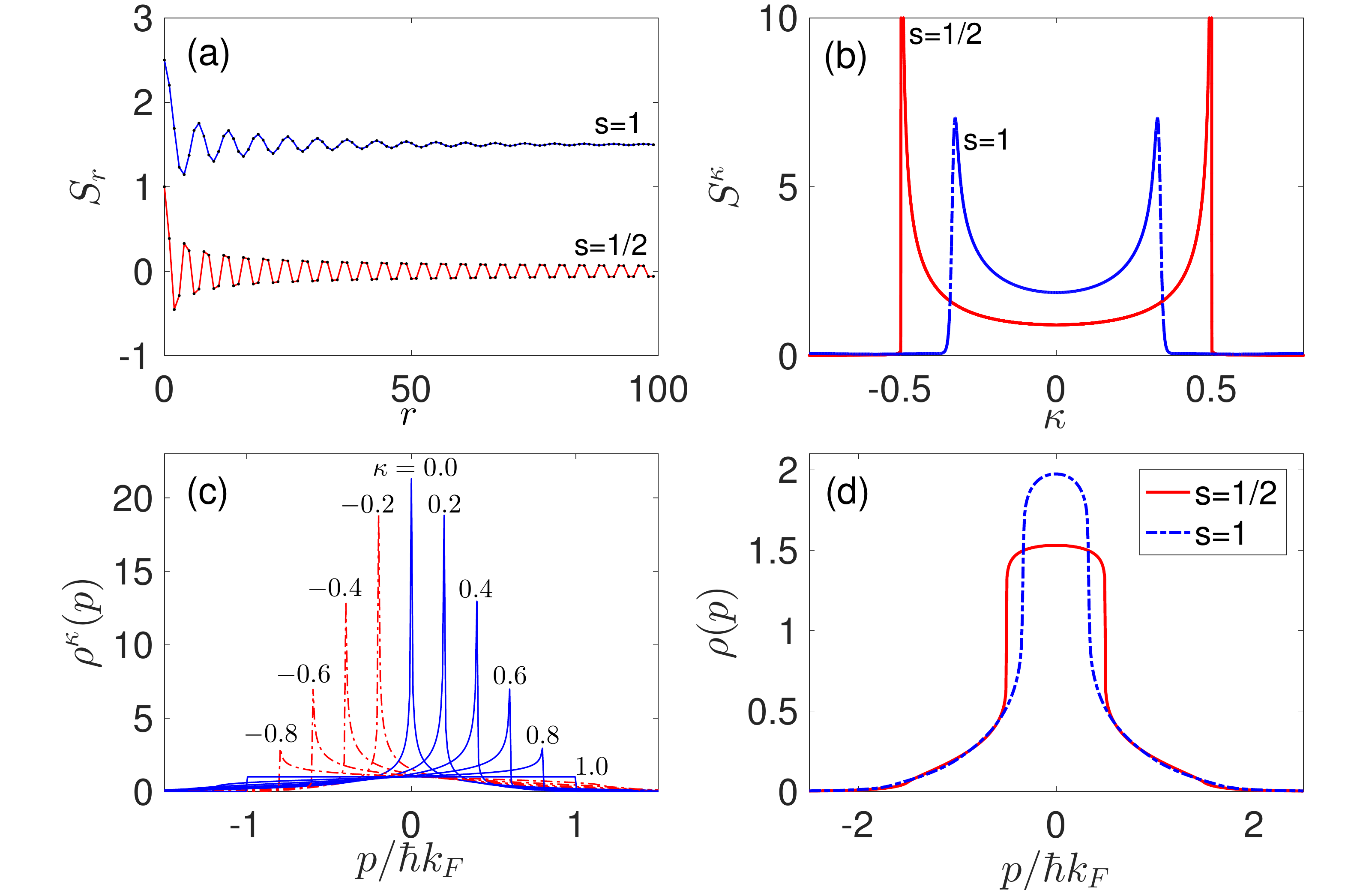}
\caption{Spin correlation function and momentum distribution of translational invariant system. {\bf (a)} $S_r$ calculated by iTEBD for an infinite chain. {\bf (b)} $S^{\kappa}$ obtained by Fourier transform of $S_r$ with $r$ up to 10000. {\bf (c)} Momentum distribution of hardcore anyon gas $\rho^{\kappa}(p)$ for $N$=201. {\bf (d)} Momentum distribution (summed over all spin components) of the spinor gases for $N$=201 particles. Figure extracted from Ref.~\cite{Yang2017}.}
\label{momentum}
\end{figure}

The OBDM for the homogeneous hardcore anyon gas, $\rho^{\kappa}(y)$, has an analytic expression in the form of the
Toeplitz determinant \cite{Calabrese2007,Santachiara2007, Santachiara2008}.
Its momentum distribution, $\rho^{\kappa}(p)$, is investigated in Ref.~\cite{Santachiara2008}.
It is shown that $\rho^{\kappa}(p)$ is peaked at $p=\kappa \hbar k_F$, where $k_F=N\pi/L$ is the Fermi momentum, for
$\kappa \in (-1,1)$.
Whereas for $\kappa=\pm 1$, the system becomes an ideal spinless Fermi gas whose momentum distribution is characterized
by the Fermi sea.
Examples of $\rho^\kappa(p)$ for $N=201$ are shown in Fig.~\ref{momentum}(c). 

As examples, we consider a spin-1/2 and a spin-1 Fermi gases with spin independent interaction with $N=201$. The corresponding spin-chain models in the strong interaction limit are the SU(2) and the SU(3) Sutherland models,
respectively~\cite{Sutherland1975}.
The spin correlation functions $S_r=\sum_{\sigma}S_r(\sigma,\sigma)$, calculated using the infinite system size TEBD (iTEBD) method~\cite{Vidal2007,Kjall2013}, and
$S^{\kappa}=\sum_{\sigma}S^{\kappa}(\sigma,\sigma)$ are plotted in Fig.~\ref{momentum}(a) and (b), respectively.
The total momentum distribution functions $\rho(p)=\sum_{\sigma}\rho_{\sigma}(p)$ for the spinor gas are shown in
Fig.~\ref{momentum}(d).

We remark that the spinor quantum gas in strongly repulsive regime has been studied within the context of spin-incoherent Luttinger
liquid~\cite{Fiete2007}, and the ground state momentum distribution for SU(2) case has been studied in
Ref.~\cite{Ogata1990,Cheianov2005,Imambekov2006}, the result in Fig.~\ref{momentum}(d) can be compared with
Fig. 3 in Ref.~\cite{Ogata1990} which is for a lattice system and for up to 32 sites with a quarter filling (note that
their definition of $k_F$ differs from ours by a factor of 2).
Here we want to mention that a sophisticated method developed in Ref.~\cite{Imambekov2006} can be used to efficiently
calculate $\rho(p)$ for homogeneous spin-1/2 fermions, but our method is more flexible and much more general as it can deal
with both bosonic and fermionic systems with arbitrary spin.

\subsection{Dynamical fermionization}
The real space density profile is given by the diagonal elements of the OBDM. The OBDM $\rho(x',x;\sigma',\sigma)$ associated with an SCAW is given by Eq.~(\ref{OBDMspincharge}). Correspondingly, the real space density profile of the strongly interacting spinor gas is given by
\begin{equation}
    n_\sigma(x) = \rho(x,x;\sigma,\sigma)=\sum_{m=1}^{N}\rho_{mm}(x,x)S_{mm}(\sigma,\sigma)\,,
\end{equation}
which depends on the spin configuration $\chi$ through the spin correlation function $S_{mm}(\sigma,\sigma)$. However, using $\sum_\sigma S_{mm}(\sigma,\sigma)=1$, one can readily show that the total density profile, summed over all spin components, is given by
\begin{equation}
    n(x) = \sum_{\sigma} n_\sigma(x) = \sum_{m=1}^{N}\rho_{mm}(x,x) = n_F(x)\,,
\end{equation}
is spin-independent and coincides with the density profile of the spinless Fermi gas $n_F(x)$. This phenomenon is sometimes called {\it fermionization}, which can be intuitively understood as resulting from the strong repulsive interaction between particles which mimics the statistical repulsion between identical fermions.
The momentum distribution of a spinor gas, by contrast, does not exhibit a similar fermionization. As Fig.~\ref{momentum}(d) shows, even the total momentum distribution depends on the spin configuration $\chi$. 

In the previous studies of spinless hardcore bosons, the phenomenon of {\it dynamical fermionization} (DF) \cite{DF1,DF2,DF3} has been discovered. This refers to the following situation: the system is initially trapped in a harmonic potential and the potential is suddenly quenched such that the cloud starts to expand. The momentum distribution of the expanded cloud asymptotically approaches that of an ideal spinless Fermi gas in the initial harmonic trap. Recently, DF has been observed in experiment \cite{DFexp}. Theoretically, a hardcore spinless anyonic gas has
also been shown to exhibit DF \cite{DFanyon}. With the tools developed above, we can now examine such a phenomenon in a strongly interacting spinor gas.

Consider a harmonically trapped spinor gas. Let us first focus on the hardcore limit. The wavefunction takes the SCAW form (\ref{scaw}), where the spin state $\chi$ is arbitrary due to the spin degeneracy in the hardcore limit, and the charge state $\varphi$ is the Slater determinant constructed from the $N$ lowest-energy single-particle harmonic
oscillator eigenstates $\phi_n(x)=(2^n n!\sqrt{\pi})^{-1/2} H_n(x)e^{-x^2/2}$ $(n=0,1,...,N-1)$, which we denote as 
\begin{equation}
    \varphi(0) = {\rm Det}[\phi_0(x),\phi_1(x),...,\phi_{N-1}(x)]/\sqrt{N!}\,.
\end{equation}
At $t=0$, the trap is suddenly turned off. Crucially, due to
the hardcore constraint, the spin configuration remains
frozen. As a consequence, the spin correlation function
$S_{mn}(\sigma',\sigma)$ in the OBDM [Eq.~(\ref{OBDMspincharge})] does not evolve in time. The time dependence of the OBDM is carried by the spatial part $\rho_{mn}(x',x)$, and hence $\varphi(t)$, according to Eq.~(\ref{SOBDM}). On the other hand, $\varphi(t)$ is related to $\varphi(0)$ as 
\begin{equation}
     \varphi(x_1,x_2,...,x_N;t)=b^{-N/2} \varphi\left(\frac{x_1}{b},\frac{x_2}{b},..., \frac{x_N}{b};0\right) 
  \exp \left[{i\left( \frac{\dot{b}}{b}\sum_i^N \frac{x_i^2}{2}-\sum_i^N E_i\tau(t)\right)} \right],
    \label{eq:scale}
\end{equation}
where $E_i$ is the energy of the $i^{\rm th}$ single-particle eigenstate of the initial harmonic trap, $b(t)=\sqrt{1+t^2}$ the spatial scaling parameter, and $\tau(t)=\int_0^t dt'/b^{2}(t')$ the temporal scaling parameter. Equation~(\ref{eq:scale}) follows from the scaling solution
of the harmonic oscillator state under a parametric modulation of the trapping frequency \cite{scaling}. With Eq.~(\ref{eq:scale}), one can readily show that the OBDM at time $t$ is also related to the initial OBDM through a scaling transformation:
\begin{equation}
    \rho(x',x;\sigma',\sigma;t)=\frac{1}{b}\exp\left[\frac{i\dot{b}}{2b}(x^2-{x'}^2)\right]\rho(x'/b,x/b;\sigma',\sigma;0).
    \label{eq:slOBDM}
\end{equation}
It follows immediately that the real space density profile at time $t$ is given by
\begin{equation}
    n_\sigma(x;t) = \rho(x,x;\sigma,\sigma;t)=\frac{1}{b}\,n_\sigma(x/b;0) \,,
\end{equation}
which describes a self-similar expansion for each spin
component.

To obtain the momentum distribution, we need to take the Fourier transform of Eq.~(\ref{eq:slOBDM}). The integral in general does not yield closed form expression. However, in the asymptotic limit $t\rightarrow \infty$ (for which $b \rightarrow t$ and $\dot{b} \rightarrow 1$), the integral can be greatly simplified by invoking the stationary phase approximation \cite{DF1} due to the fast oscillating nature of the integrand, and we obtain
\begin{equation}
    \rho_\sigma (p;t \rightarrow \infty) = \rho(k,k;\sigma,\sigma;0) = n_\sigma(p;0)\,
    \label{dfspin}
\end{equation}
which means the asymptotic momentum distribution of of the spin-$\sigma$ component has the same shape as the initial real space density profile inside the trap.  It is amusing to
note that this is just the opposite situation of the ballistic
expansion under which the asymptotic real space density
profile takes the shape of the initial momentum distribution
in the trap. The total momentum distribution therefore has the property
\begin{equation}
    \rho(p;t \rightarrow \infty) = \sum_\sigma \rho_\sigma (p;t \rightarrow \infty) = n_F(p;0)\,,
    \label{dftotal}
\end{equation}
and therefore takes the shape of the initial total real space
density profile, which is the same as the momentum
distribution $\rho_F(p)$ of the spinless Fermi gas in the trap. Equations (\ref{dfspin}) and (\ref{dftotal}) sum up the properties of DF for a hardcore spinor gas \cite{DFspinor}.

Now let us consider the case where the interaction strength is large but finite. The discussion above on the hardcore case relies on the fact that the spin degrees of freedom is frozen for hardcore particles. It may seem that, away from the hardcore limit, DF should not occur since now the spin degrees of freedom is released and governed by the effective spin-chain Hamiltonian $H_{\rm sc}$ which becomes time-dependent after the quench of the trapping potential. However, as one can easily see, in this case the coefficients $C_i$ in $H_{\rm sc}$ have the scaling behavior as $C_i(t) = C_i(0)/b^3(t)$. As a result, we have \cite{Volosniev2016}
\begin{equation}
    H_{\rm sc}(t) = \frac{1}{b^3(t)}\,H_{\rm sc}(0)\,, \label{scalingH}
\end{equation}
which means that an eigenstate of the initial spin-chain Hamiltonian $H_{\rm sc}(0)$ remains as an eigenstate of $H_{\rm sc}(t)$ for $t>0$. In this way, the spin degrees of freedom is effectively frozen, just as in the hardcore case. Therefore, all the DF properties obtained for hardcore spinor gas remains valid for large but finite interaction strength. We emphasize that the scaling behavior of the spin-chain Hamiltonian, Eq.~(\ref{scalingH}), is a special property for harmonic traps. For the quench of an initial trapping potential that is non-harmonic, we do not expect this to be the case and hence DF should not occur under such a situation. 

\section{Conclusion}
In this article, we provided a short review of our work on strongly interacting spinor quantum gases. Through a generalized Bose-Fermi mapping, we are able to map the strongly interacting system into a weakly interacting one, whose charge degrees of freedom is described by a spinless Fermi gas while the spin degrees of freedom by an effective spin-chain Hamiltonian derived from a perturbative approach. The wavefunction of the system takes the form of the SCAW, which takes the spin-charge separated form in a given spatial sector. This allows us to calculate certain collective excitation frequencies, the OBDM, as well as the momentum distribution in an efficient way. Finally, we discussed the dynamical fermionization of the spinor gas in an initially harmonic trap that is suddenly quenched, and show that the asymptotic momentum distribution is intimately connected to the initial real space density profile. This represents a rare case where exact results can be obtained for a many-body system.

\ack We would like to than Prof. Xiwen Guan for many insightful discussions, and Liming Guan and Timothy Skaras for their contribution to part of the original work. This work was supported by the US NSF and the Welch Foundation (Grant No. C-1669).

\section*{References}
  \bibliographystyle{iopart-num}
    \bibliography{1D_bibliography}
\end{document}